\documentclass[sn-mathphys,Numbered]{sn-jnl}

\usepackage{graphicx}%
\usepackage{multirow}%
\usepackage{amsmath,amssymb,amsfonts}%
\usepackage{amsthm}%
\usepackage{mathrsfs}%
\usepackage[title]{appendix}%
\usepackage[dvipsnames]{xcolor}
\usepackage{textcomp}%
\usepackage{manyfoot}%
\usepackage{booktabs}%
\usepackage{algorithm}%
\usepackage{algorithmicx}%
\usepackage{algpseudocode}%
\usepackage{listings}%
\usepackage{caption}%
\usepackage{subcaption}%
\usepackage{cleveref}%
\usepackage{multirow}
\usepackage{booktabs}\let\cline\cmidrule
\usepackage{cellspace}
\theoremstyle{thmstyleone}%
\theoremstyle{thmstyletwo}%

\theoremstyle{thmstylethree}%

\begin{document}

\title[Effects of Variable Mass, Disk-Like Structure, and Radiation Pressure on the Dynamics of Circular Restricted Three-Body Problem]{Effects of Variable Mass, Disk-Like Structure, and Radiation Pressure on the Dynamics of Circular Restricted Three-Body Problem}


\author[1]{\fnm{L. B.} \sur{Putra}}

\author*[2]{\fnm{I.} \sur{Nurul Huda}}\email{ibnu.nurul.huda@brin.go.id}

\author[1]{\fnm{H. S.} \sur{Ramadhan}}

\author[1,2]{\fnm{M. B.} \sur{Saputra}}

\author[3]{\fnm{T.} \sur{Hidayat}}

\affil[1]{\orgdiv{Departemen Fisika}, \orgname{FMIPA, Universitas Indonesia}, \orgaddress{ \city{Depok, 16424},  \country{Indonesia}}}

\affil*[2]{\orgdiv{Research Center for Computing}, \orgname{National Research and Innovation Agency}, \orgaddress{ \city{Bogor},  \country{Indonesia}}}

\affil[3]{\orgdiv{Department of Astronomy and Bosscha Observatory}, \orgname{Institut Teknologi Bandung}, \orgaddress{\city{Bandung}, \country{Indonesia}}}


\abstract{In this paper, we intend to investigate the dynamics of the Circular Restricted Three-Body Problem. Here we assumed the primaries as the source of radiation and have variable mass. The gravitational perturbation from disk-like structure are also considered in this study. There exist five equilibrium points in this system. By considering the combined effect from disk-like structure and the mass transfer, we found that the classical collinear equilibrium points depart from x-axis. Meanwhile, this combined effect also breaks the symmetry of tringular equlibrium point positions. We noted that the quasi-equilibrium points are unstable whereas the triangular equilibrium points are stable if the mass ratio $\mu$ smaller than critical mass $\mu_c$. It shows that the stability of triangular equilibrium points depends on time.}




\maketitle

\section{Introduction}\label{sec1}

Circular Restricted Three Body Problem (CRTBP) consists of the movement of the third body with respect to the two primaries. The primaries move in a circular orbit and the third body is influenced by but not influences the primaries. In the classical case, the primaries and the third body are assumed as the point mass \citep[see e.g.][]{roy2004orbital,murray2000solar}. There exist five equilibrium points which are divided into two categories named collinear equilibrium points $L_{1}$, $L_{2}$, and $L_{3}$ and triangular equilibrium points $L_{4}$ and $L_{5}$. The collinear equilibrium points are always unstable while the triangular equilibrium points are stable if the mass ratio $\mu < \mu_c = 0.038520896504551$.

The complexity of nature has made the CRTBP not suitable for some cases. Therefore, some authors have tried to develop the CRTBP by incorporating various effects. For instance, \citet{radzievskii1950restricted} and \citet{chernikov1970photogravitational} have considered the effect of photogravitation in the CRTBP for mimicking the stellar objects . More recently, the influence of disk-like structure has been incorporated in the CRTBP \citep[see e.g.][]{chermnykh1987stability,jiang2004chaotic}. There are several studies that combined various additional effects in CRTBP. For instance, \citet{SinghTauraGCRTBP} has studied CRTBP by assuming both primaries are radiating and oblate bodies, together with the effect of disk-like structure. \citet{huda2023studying} combined the effect of photogravitational and disk-like structure, with addition of oblateness and finite-straight segment for the primaries, to study the stability of equlibrium points in CRTBP.

Several close binary star systems have been discovered \citep{tutukov2020evolution,price2020close}. It is already studied that some of them have planets that have mass much less compared to the binary \citep{gong2018formation,thebault2015planet}. Meanwhile, previous studies also suggest that there is also possibility that an asteroid belt like structure also exist in the binary system \citep{bancelin2015asteroid,jennings2020pulsar}. In certain cases, the transfer of mass between binary stars is unavoidable \citep{qian2020contact}. However, accurately predicting how mass moves between close-orbiting stars is still a major challenge. In the case of CRTBP, the transfer mass between star in binary sytem can be modelled by the variability of mass of each primary. The study of variable mass in the restricted three-body problem was done by \citet{orlov1939existence} in 1930s. More recently, \citet{Lukyanov2005ConservativeTP} studied the CRTBP system which the primaries have variable masses but the sum of their masses remains constant. \citet{singh2012equilibrium} consider the variation of mass of primaries in accordance with the combined Meshcherskii law. Several studies also consider the variable mass of the third body \citep[see e.g.][]{albidah2023shapes,suraj2021modified,abouelmagd2015out}.


In this study, we investigate the possible movement of the infinitesimal mass in the close binary star system. We used a framework of CRTBP where the binaries are primaries. We assumed that the stars emit radiation and there is a mass transfer between primaries. We also considered a disk-like structure surrounding this three-body system, mimicking the Kuiper or asteroid belt structure. 

This paper is outlined as follows. In the Section \ref{sec2}, we give a detail about the equation of motion of the system. The detail about the equilibrium points is given in Section \ref{sec3}. Section \ref{sec4} describes the stability of the system. Finally, the conclusion is given in Section \ref{sec6}. Here we used Mathematica software to conduct a numerical calculation or algebraic manipulation.

\section{Equation of Motion}\label{sec2}

Let the mass of the first and second primaries are $m_1$ and $m_2$ respectively. The mass ratio between primaries is represented by $\mu = m_2/(m_1 + m_2)$, where $0 < \mu < 1$. Hence we represent the mass of the primaries by $1-\mu$ and $\mu$. For simplifying the problem, we consider the system in a two dimensional rotational coordinate $Oxy$ and the primaries always lays on $x$-axis. The origin of the coordinate system is located in the positon of $m_1$. We take the distance between primaries as the unit of length and the unit of time is chosen in such a way so that the gravitational constant is unity. Let ($x$, $y$) be the position of the third body. We follow \citet{Lukyanov2005ConservativeTP} for describing the CRTBP where the primaries have variable mass. However, we also consider the effect of the radiation pressure on the primaries and the disk-like structure surrounding the three-body system. 

The radiation force ($F_p$) has an opposite direction with respect to the gravitational force ($F_g$). In order to consider the radiation pressure in the CRTBP, we defined the mass reduction factor $q = 1 - (F_p/F_g)$, where $0 < 1 - q \ll 1$. Meanwhile, the disk-like structure effect can be modelled by following \citet{miyamoto1975three}. The potential of disk-like structure for planar version is given as
\begin{equation}
    V(x,y) = \frac{M_b}{\sqrt{r^2 + T^2}}
\end{equation}
where $M_b$ is the total mass of the disk-like structure and $r^2 = x^2 + y^2$ is the radial distance of the infinitesimal mass. The mass parameter of the disk-like structure is $M_b$ and is assumed to be small compared to the total mass of the primaries ($M_b<<1$). Here $T = a + b$ means the dust belt's density profile, where $a$ and $b$ are the flatness and core parameters of the disc respectively. 

\begin{figure}
         \centering
         \includegraphics[width=0.5\textwidth]{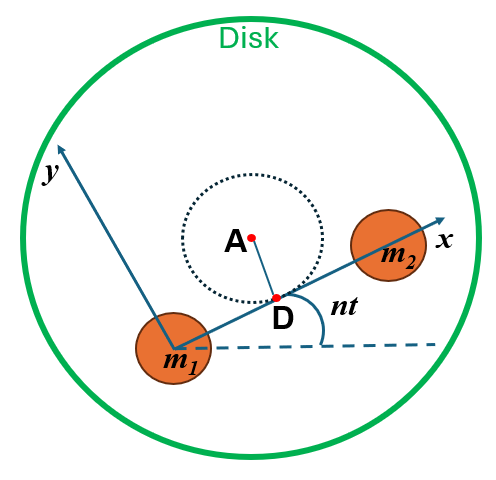}
     \caption{Schematic diagram of the system in this study.}
        \label{fig:schematic}
\end{figure}

Figure \ref{fig:schematic} shows a graphic representation of the system. We follow \cite{Lukyanov2005ConservativeTP} to model the mass transfer between the primaries. It has a center of inertia ($A$) as its center. This center is a static point. There is also center of mass ($D$) that moves around point $A$. It has to be noted that in our case the distance between $A$ and $D$ is so small. For simplicity, we shall consider conservative linear mass transfer law between the primaries,
\begin{equation}
\label{mut}
    \mu(t)=\frac{m_2(t)}{m_1(t)+m_2(t)}=kt.
\end{equation}
Here $t$ means time and $0<t<\frac{1}{k}$. It has to be noted that the sum of mass $m_1(t)$ and $m_2(t)$ is constant. We assume that the rate of transfer $k$ is much slower compared to the orbital period of the primaries, i.e. $k<<\frac{1}{n}$ where $n$ is the mean motion of the two body system,
\begin{equation}
    n^2=1+\frac{2M_br_c}{(r_c^2+T^2)^{3/2}}.
\end{equation}
The reference radius of the disk-like structure is given by $r_c^2=1-\mu+\mu^2$ as in \cite{SinghTauraGCRTBP}. Assuming that the transfer mass between primaries is very slow and that the dominant order in the expansion is the first order, we have
\begin{equation}
    \mu (t)\approx\mu (t_0)+\Dot{\mu}(t_0)(t-t_0),
\end{equation}
where we could define $\Dot{\mu}(t_0)\equiv k$ and $\mu(t_0)\equiv\mu_0$, so that
\begin{equation}
\label{mutv2}
    \mu (t)=\mu_0+k(t-t_0).
\end{equation}
Note that for $\mu_0=kt_0$, \cref{mutv2} reverts back to \cref{mut}. In the case of $\mu_0=kt_0$, the domain for $t$ is $\left(t_0-\frac{\mu_0}{k}\right)<t<\left(t_0+\frac{1-\mu_0}{k}\right)$.

The equation of motion of the system is given as follows
\begin{align} \label{eq:eq_motion}
\begin{split}
    \Ddot{x}&-2n\Dot{y}=W_x, \\
    \Ddot{y}&+2n\Dot{x}=W_y,
\end{split}
\end{align}
where $W_x$ and $W_y$ mean the partial derivative of $W$ with respect to $x$ and $y$ respectively. The pseudo potential is given by
\begin{equation}
    W=\frac{1}{2}n^2(x^2+y^2)-\mu n^2 x+\frac{(1-\mu)q_1}{r_1}+\frac{\mu q_2}{r_2}+\frac{M_b}{(\mathcal{R}^2+T^2)^{1/2}}.
\end{equation}
The first derivatives of the pseudo potential with respect to the third body position is given by
\begin{align}
\begin{split}
    W_x&=n^2x-\mu n^2-\frac{(1-\mu)q_1x}{r_1^3}-\frac{\mu q_2(x-1)}{r_2^3}-\frac{M_b(x-\mu)}{(\mathcal{R}^2+T^2)^{3/2}}, \\
    W_y&=n^2y-\frac{(1-\mu)q_1y}{r_1^3}-\frac{\mu q_2y}{r_2^3}-\frac{M_b(y-2k/n)}{(\mathcal{R}^2+T^2)^{3/2}}.
\end{split}
\end{align}
Here $q_1$ and $q_2$ are the radiation pressure factor for $m_1$ and $m_2$. If we do not consider the radiation and disk-like structure effects, \cref{mut} will be similar to the equation of motion in \cite{CRTBPLukyanov}. We consider the same coordinate system in \cite{CRTBPLukyanov} where the origin of the rotational coordinate is the position of $m_1$, hence
\begin{align} \label{eq:r1r2_0}
\begin{split}
    r_1^2&=x^2+y^2, \\
    r_2^2&=(x-1)^2+y^2.
\end{split}
\end{align}
Since the center of the disk-like structure is the point $C$, i.e. the point around which the primaries barycenter orbits, the radial distance of the infinitesimal mass becomes
\begin{equation}
    \mathcal{R}^2=(x-\mu)^2+(y-2k/n)^2.
\end{equation}

\section{Equilibrium Points}\label{sec3}

\subsection{Quasi-Collinear Points}

The collinear points $L_1$, $L_2$, and $L_3$ are the solution located in the interval $1 < x < \infty$, $0 < x < 1$, and $-\infty < x < 0$, respectively. The position of collinear points are found by considering $\Ddot{x} = \Ddot{y} = \dot{x} = \dot{y} = y = 0$ into \cref{eq:eq_motion}. We have
\begin{align} \label{eq:colpotensial}
\begin{split}
    (x-\mu)\left(n^2-\frac{M_b}{((x-\mu)^2+\frac{4k^2}{n^2}+T^2)^{3/2}}\right)-\frac{(1-\mu)q_1x}{x^3}-\frac{\mu q_2(x-1)}{(x-1)^3}=0, \\
    \frac{2M_bk}{n((x-\mu)^2+\frac{4k^2}{n^2}+T^2)^{3/2}}=0.
\end{split}
\end{align}
It is clear from \cref{eq:colpotensial} that the collinear equilibrium points only exist if $M_b = 0$ or $k = 0$. Nevertheless, we searched a possible equilibrium points near x-axis when $M_b \neq 0$ and $k \neq 0$ by calculating the equilibrium points numerically. The numerical values are obtained using a numerical algorithm in Mathematica. Here we consider $\mu_0=0.3$ and $t_0=0$. The resulting time dependence graph can be seen in \cref{fig:L12345M,fig:L12345K,fig:L12345Q}. As expected, we found that the equilibrium points are quasi-collinear. They shifted slightly towards the $+y$ axis, due to the existence of mass variation and the disk like structure, which has the point $C$ (that is not on the barycenter, nor is it anywhere in the $x$ axis) as its center.

According to \cref{fig:L1M,fig:L2M,fig:L3M}, the position of $L_1$, $L_2$, $L_3$ have affected by $M_b$. Higher $M_b$ makes $L_1$, $L_2$, and $L_3$ position further away from x-axis. Meanwhile, higher $k$ means that the position of $L_1$, $L_2$, and $L_3$ are shifted higher with respect to the original position as time increase (see \cref{fig:L1K,fig:L2K,fig:L3K}). In contrast, as shown in \cref{fig:L1Q,fig:L2Q,fig:L3Q} the value of $q_1$ and $q_2$ has contributed lower compared to $k$ in determining the shift of $L_1$, $L_2$, and $L_3$ as time increase.


\subsection{Triangular Points}

In order to find the position of equilibrium points, we have to solve \cref{eq:eq_motion} by considering $\Ddot{x} = \Ddot{y} = \dot{x} = \dot{y} = 0$. The position of  triangular equilibrium points can be calculated by considering $y \neq 0$. Hence we get
\begin{align} \label{eq:tripotensial}
\begin{split}
    (x-\mu)\left[n^2-\frac{M_b}{(\mathcal{R}^2+T^2)^{3/2}}\right]-\frac{(1-\mu)q_1x}{r_1^3}-\frac{\mu q_2(x-1)}{r_2^3} = 0, \\
    y \left[n^2-\frac{M_b}{(\mathcal{R}^2+T^2)^{3/2}}\right]   -\frac{(1-\mu)q_1y}{r_1^3}-\frac{\mu q_2y}{r_2^3}+\frac{2kM_b}{n(\mathcal{R}^2+T^2)^{3/2}} = 0.
\end{split}
\end{align}
We assume that the position of triangular points in the modified CRTBP is the perturbed version of classical case ($r_1 = 1; r_2 = 1$), i.e.
\begin{align} \label{eq:r1r2}
\begin{split}
    r_1&=1+\epsilon_1, \\
    r_2&=1+\epsilon_2,
\end{split}
\end{align}
where $\epsilon_{1,2} \ll 1$. By substituting \cref{eq:r1r2} to \cref{eq:r1r2_0}, neglecting higher order of $\epsilon_{1,2}$, and solving for $x$ and $y$, we have position of triangular equilibrium points as follows
\begin{align} \label{eq:triangularpoint1}
    \begin{split}
        x &= \frac{1}{2}+\epsilon_1-\epsilon_2, \\
        y &= \sqrt{\frac{3}{4}+\epsilon_1+\epsilon_2},
    \end{split}
\end{align}
Following \cite{SinghTauraGCRTBP}, we consider \cref{eq:r1r2} and \cref{eq:triangularpoint1} into \cref{eq:tripotensial}. Hence, with additional expansion of $k$ to the first order, we obtain
\begin{align} \label{eq:epsilon}
\begin{split}
    \epsilon_1&=-\frac{1-q_1}{3}+\frac{M_b(1-2r_c)}{3(r_c^2+T^2)^{3/2}}, \\
    \epsilon_2&=-\frac{1-q_2}{3}+\frac{M_b(1-2r_c)}{3(r_c^2+T^2)^{3/2}}.
\end{split}
\end{align}
Substituting \cref{eq:epsilon} to \cref{eq:triangularpoint1} yields the triangular points $L_4$ and $L_5$
\begin{equation} 
\label{xTri}
    x=\frac{1}{2}-\frac{q_2-q_1}{3}
\end{equation}
and
\begin{equation}
\label{yTri}
    y=\pm\frac{\sqrt{3}}{2}\left(1-\frac{2}{9}(2-q_1-q_2)+\frac{4}{9}\frac{M_b(1-2r_c)}{\left(r_c^2+T^2\right)^{3/2}}\right).
\end{equation}
It can be seen that the triangular points for this system are identical (to the first order) with the constant primary mass counterpart, albeit with a (slow) time dependence.

\begin{figure}
     \centering
     \begin{subfigure}[b]{\textwidth}
         \centering
         \includegraphics[width=0.33\textwidth]{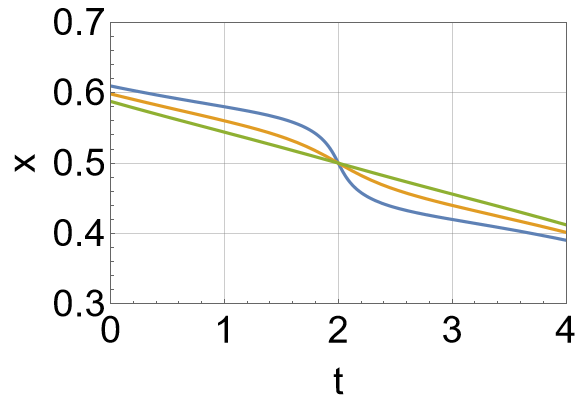}
        \includegraphics[width=0.33\textwidth]{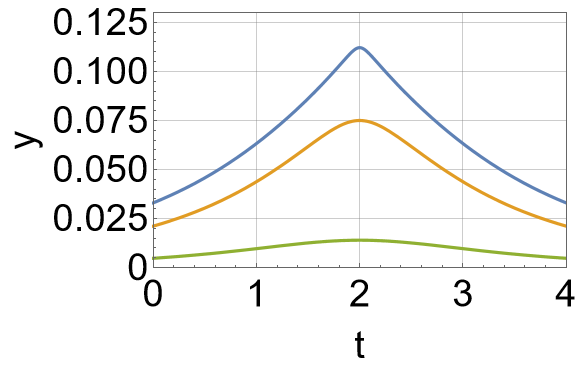}
         \caption{$L_1$}
         \label{fig:L1M}
     \end{subfigure}
     \begin{subfigure}[b]{\textwidth}
         \centering
         \includegraphics[width=0.33\textwidth]{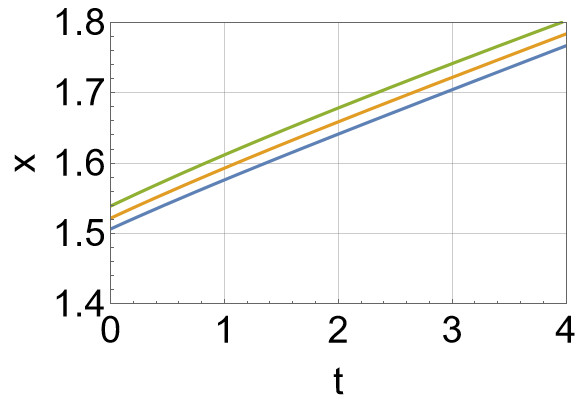}
        \includegraphics[width=0.33\textwidth]{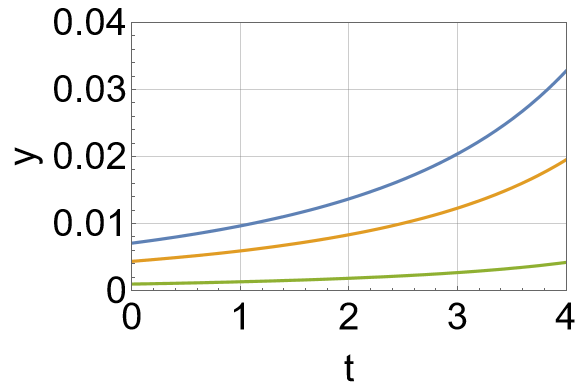}
         \caption{$L_2$}
         \label{fig:L2M}
     \end{subfigure}
     \begin{subfigure}[b]{\textwidth}
         \centering
         \includegraphics[width=0.33\textwidth]{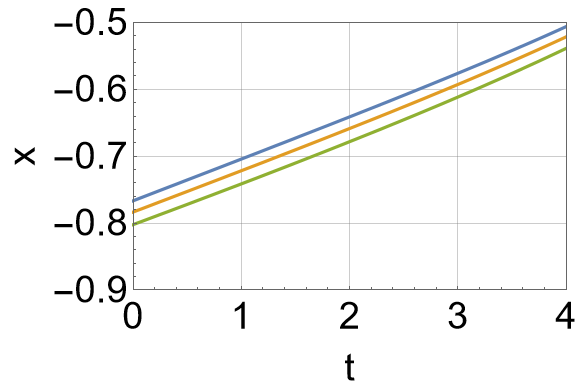}
        \includegraphics[width=0.33\textwidth]{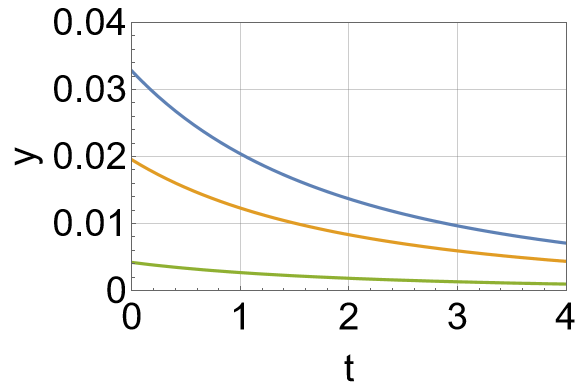}
         \caption{$L_3$}
         \label{fig:L3M}
     \end{subfigure}
     \begin{subfigure}[b]{\textwidth}
         \centering
         \includegraphics[width=0.33\textwidth]{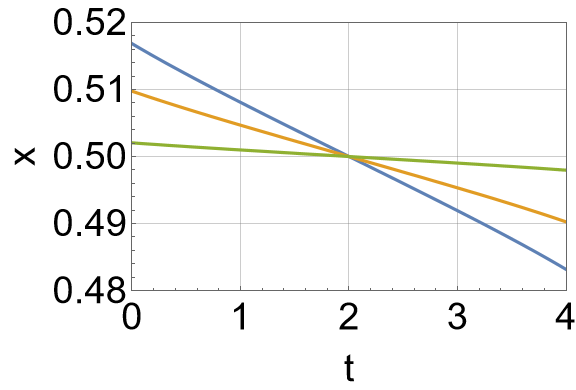}
        \includegraphics[width=0.33\textwidth]{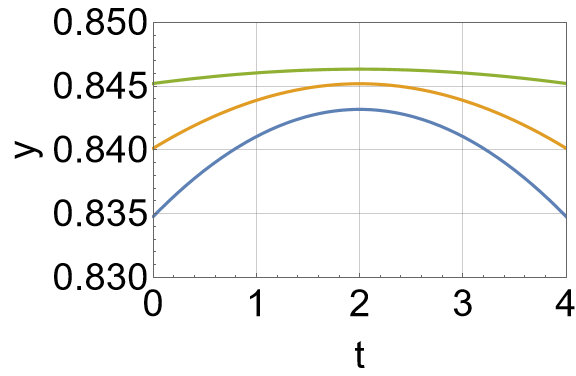}
         \caption{$L_4$}
         \label{fig:L4M}
     \end{subfigure}
     \begin{subfigure}[b]{\textwidth}
         \centering
         \includegraphics[width=0.33\textwidth]{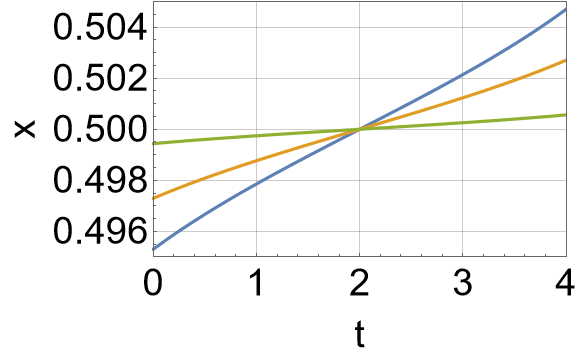}
        \includegraphics[width=0.33\textwidth]{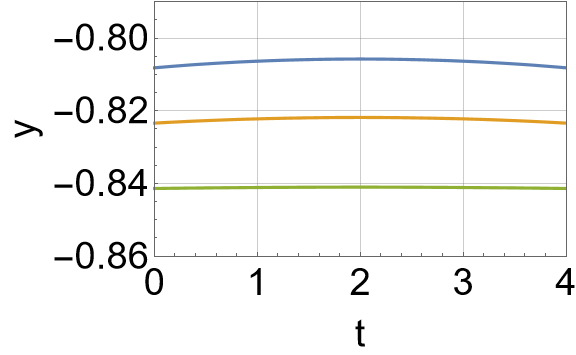}
         \caption{$L_5$}
         \label{fig:L5M}
     \end{subfigure}   
        \caption{The position of equilibrium points for $M_b = 0.09$ (\textcolor{NavyBlue}{$\blacksquare$}), $M_b = 0.05$ (\textcolor{Dandelion}{$\blacksquare$}), $M_b = 0.01$ (\textcolor{YellowGreen}{$\blacksquare$}). Here $k = 0.1$ and $q_1 = q_2 = 0.95$. We assumed $\mu_0=0.3$ and $t_0=0$.}
        \label{fig:L12345M}
\end{figure}

\begin{figure}
     \centering
     \begin{subfigure}[b]{\textwidth}
         \centering
         \includegraphics[width=0.33\textwidth]{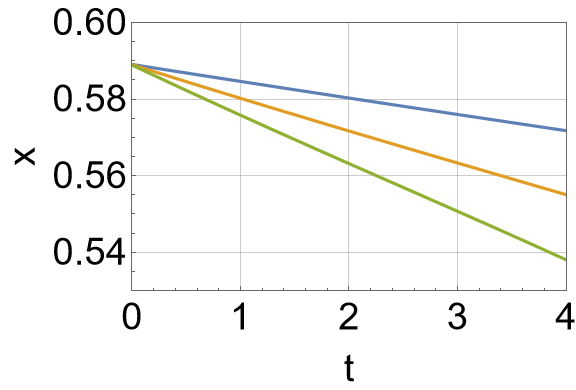}
        \includegraphics[width=0.33\textwidth]{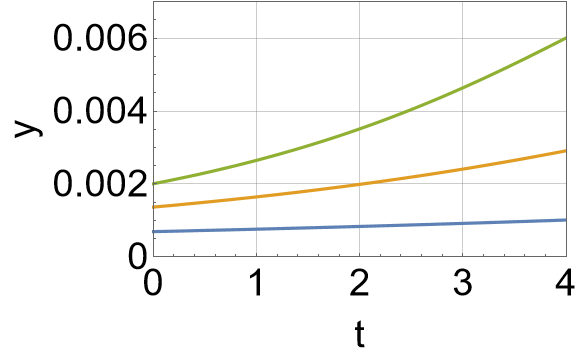}
         \caption{$L_1$}
         \label{fig:L1K}
     \end{subfigure}
     \begin{subfigure}[b]{\textwidth}
         \centering
         \includegraphics[width=0.33\textwidth]{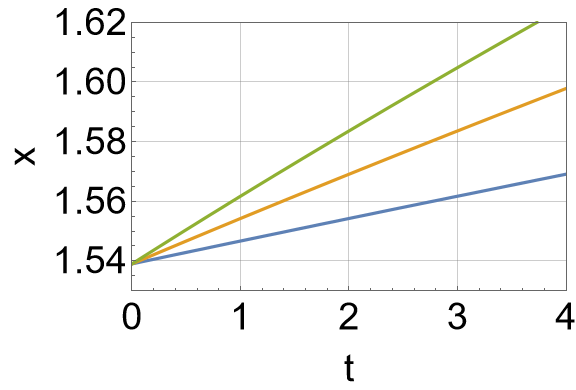}
        \includegraphics[width=0.33\textwidth]{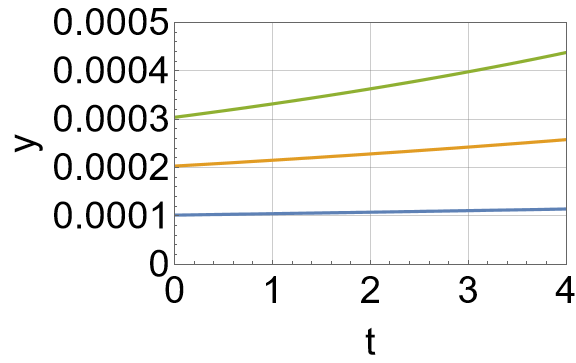}
         \caption{$L_2$}
         \label{fig:L2K}
     \end{subfigure}
     \begin{subfigure}[b]{\textwidth}
         \centering
         \includegraphics[width=0.33\textwidth]{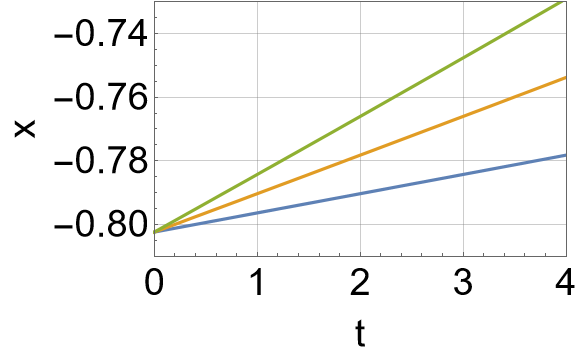}
        \includegraphics[width=0.33\textwidth]{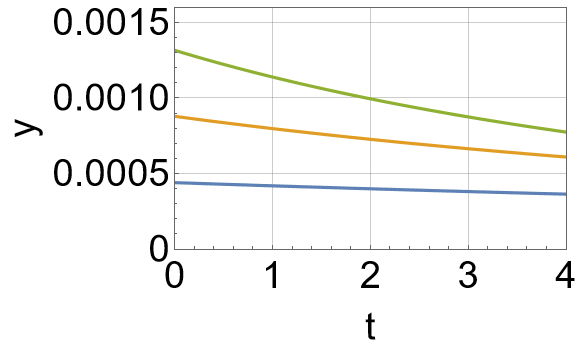}
         \caption{$L_3$}
         \label{fig:L3K}
     \end{subfigure}
     \begin{subfigure}[b]{\textwidth}
         \centering
         \includegraphics[width=0.33\textwidth]{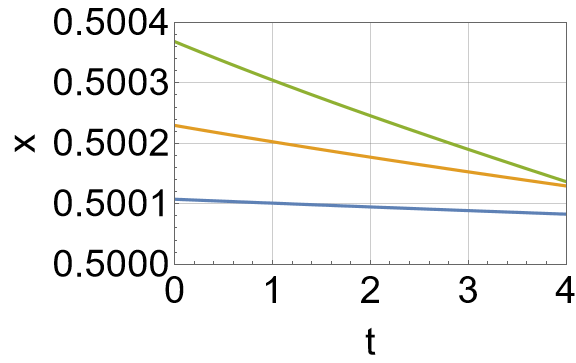}
        \includegraphics[width=0.33\textwidth]{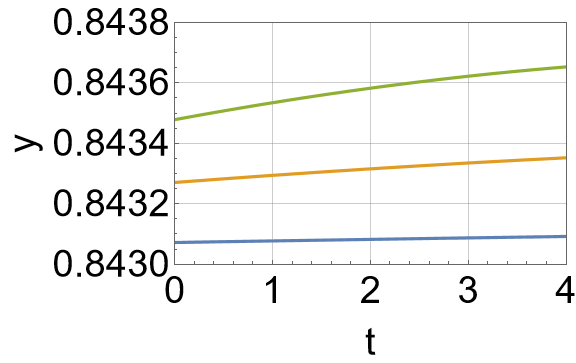}
         \caption{$L_4$}
         \label{fig:L4K}
     \end{subfigure}
     \begin{subfigure}[b]{\textwidth}
         \centering
         \includegraphics[width=0.33\textwidth]{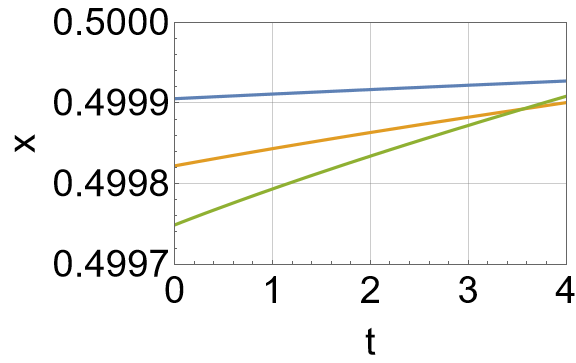}
        \includegraphics[width=0.33\textwidth]{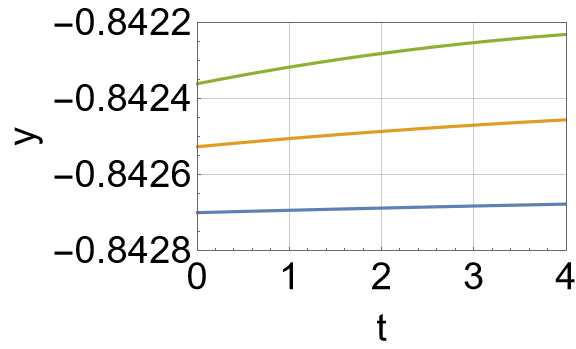}
         \caption{$L_5$}
         \label{fig:L5K}
     \end{subfigure}     
        \caption{The position of equilibrium points for $k = 0.01$ (\textcolor{NavyBlue}{$\blacksquare$}), $k = 0.02$ (\textcolor{Dandelion}{$\blacksquare$}), $k = 0.03$ (\textcolor{YellowGreen}{$\blacksquare$}). Here $M_b = 0.01$ and $q_1 = q_2 = 0.95$. We assumed $\mu_0=0.3$ and $t_0=0$.}
        \label{fig:L12345K}
\end{figure}

\begin{figure}
     \centering
     \begin{subfigure}[b]{\textwidth}
         \centering
         \includegraphics[width=0.33\textwidth]{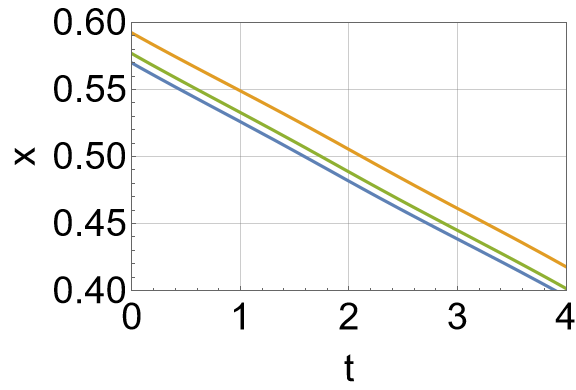}
        \includegraphics[width=0.33\textwidth]{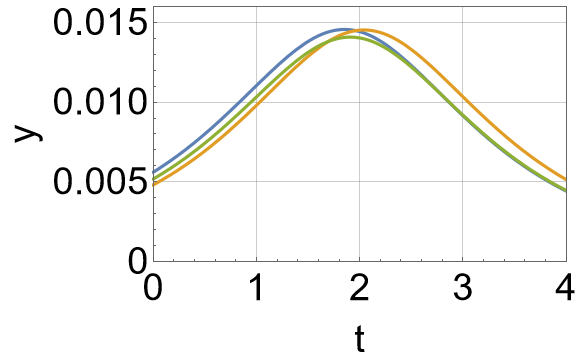}
         \caption{$L_1$}
         \label{fig:L1Q}
     \end{subfigure}
     \begin{subfigure}[b]{\textwidth}
         \centering
         \includegraphics[width=0.33\textwidth]{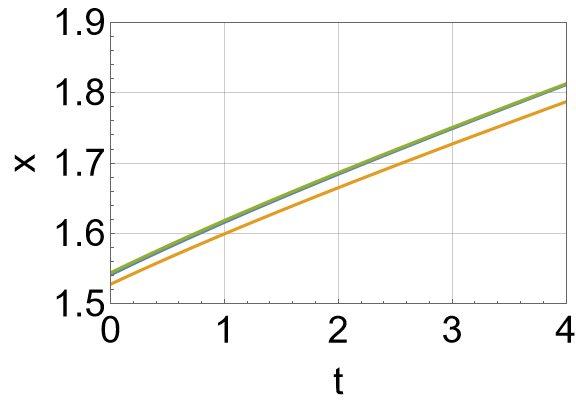}
        \includegraphics[width=0.33\textwidth]{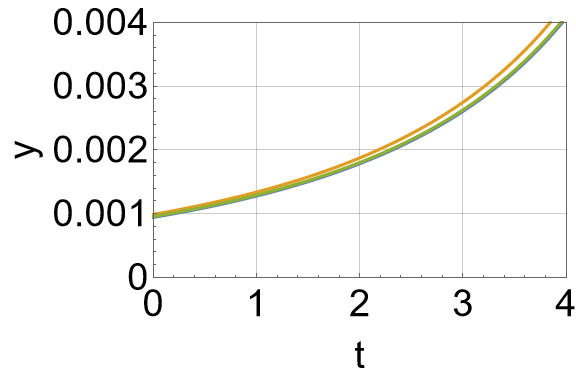}
         \caption{$L_2$}
         \label{fig:L2Q}
     \end{subfigure}
     \begin{subfigure}[b]{\textwidth}
         \centering
         \includegraphics[width=0.33\textwidth]{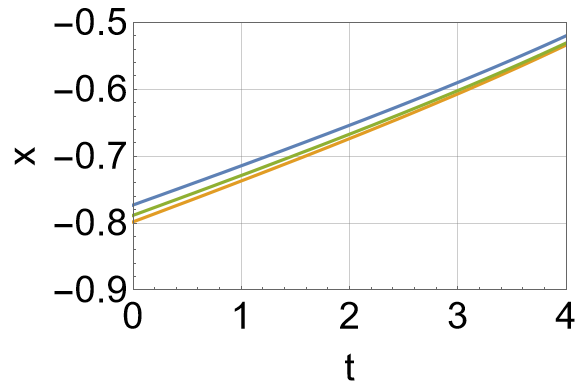}
        \includegraphics[width=0.33\textwidth]{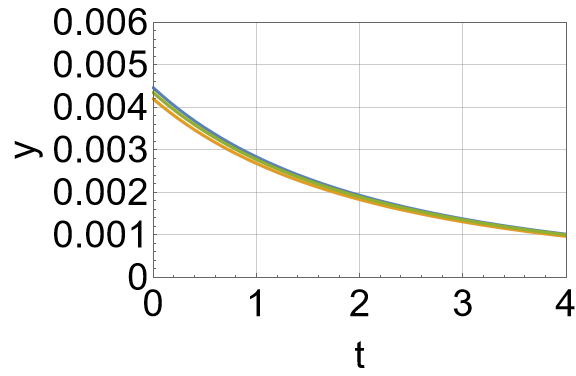}
         \caption{$L_3$}
         \label{fig:L3Q}
     \end{subfigure}
     \begin{subfigure}[b]{\textwidth}
         \centering
         \includegraphics[width=0.33\textwidth]{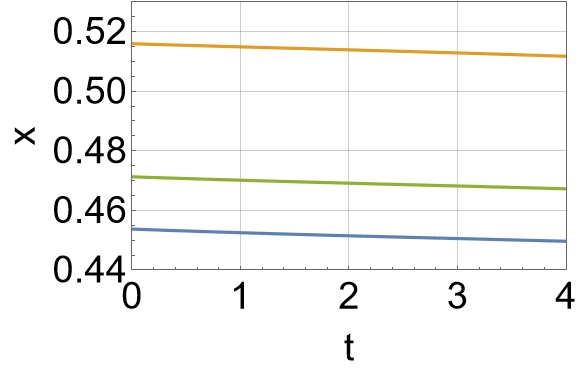}
        \includegraphics[width=0.33\textwidth]{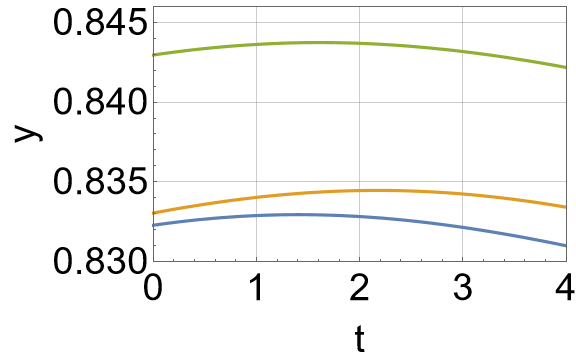}
         \caption{$L_4$}
         \label{fig:L4Q}
     \end{subfigure}
     \begin{subfigure}[b]{\textwidth}
         \centering
         \includegraphics[width=0.33\textwidth]{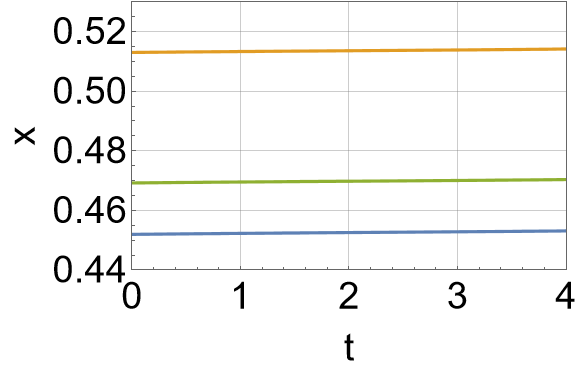}
        \includegraphics[width=0.33\textwidth]{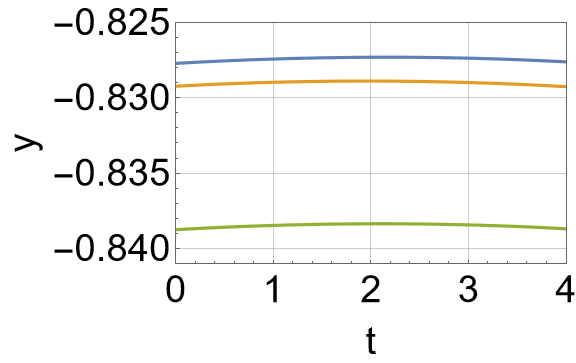}
         \caption{$L_5$}
         \label{fig:L5Q}
     \end{subfigure}     
        \caption{The position of equilibrium points for $q_1 = 0.85$; $q_2 = 0.99$ (\textcolor{NavyBlue}{$\blacksquare$}), $q_1 = 0.94$; $q_2 = 0.9$ (\textcolor{Dandelion}{$\blacksquare$}), $q_1 = 0.9$; $q_2 = 0.99$ (\textcolor{YellowGreen}{$\blacksquare$}). Here $M_b = 0.01$ and $k = 0.1$. We assumed $\mu_0=0.3$ and $t_0=0$.}
        \label{fig:L12345Q}
\end{figure}

\Cref{fig:L4M,fig:L5M} show the effect of $M_b$ in the position of triangular points. We observe that the position of $L_4$ and $L_5$ is not symmetric due to the combination of disk-like structure and mass transfer. This asymmetric is larger when $M_b$ is higher. We noted also that the position of $L_4$ and $L5$ is closer to the primaries  with increasing $M_b$. In \Cref{fig:L4K,fig:L5K}, it is clear that the decreasing value of $k$ makes $L_4$ closer to the primaries, in contrast with $L_5$. From \Cref{fig:L4Q,fig:L5Q} we observe that radiation pressure has the impact on the location of triangular equilibrium points. The location of $L_4$ and $L_5$ are closer to the source of radiation pressure when the radiation pressure getting stronger, either for $m_1$ or $m_2$.


\section{Linear Stability}\label{sec4}

\begin{table}
\caption{Characteristic roots of collinear equilibrium points with $\mu = 0.02$. We used $T = 0.2$ and $t_0 = 0$. Here $i$ means $\sqrt{-1}$. $\lambda_{2}$ and $\lambda_{4}$ have the inverse sign of $\lambda_{1}$ and $\lambda_{3}$ respectively.}
\begin{center}
{\renewcommand{\arraystretch}{1.5}
\setlength{\tabcolsep}{6pt} 
\begin{tabular}{|c|c|c|c|c|c|c|c|c|c|c|c|}
\hline
\multirow{2}{*}{Case} & \multirow{2}{*}{$1-q_1$}   & \multirow{2}{*}{$1-q_2$}   & \multirow{2}{*}{$M_b$}   & \multirow{2}{*}{$k$} & \multirow{2}{*}{$t$} & \multicolumn{2}{|c|}{$L_1$}&\multicolumn{2}{|c|}{$L_2$}&\multicolumn{2}{|c|}{$L_3$} \\ 
\cline{7-8} \cline{9-10} \cline{11-12}
&&&&& & $\lambda_{1}$ & $\lambda_{3}$ & $\lambda_{1}$ & $\lambda_{3}$ & $\lambda_{1}$ & $\lambda_{3}$\\
\hline
1&1          & 1              & 0                      & 0.1                    & 0                   & $3.0140$ & $2.3861i$ & $2.0987$ & $1.8277i$ & $0.2277$ & $1.0169i$  \\
&1          & 1              & 0                      & 0.1                    & 0.2                   & $3.1535$ & $2.4749i$ & $1.9959$ & $1.7685i$ & $0.3204$ & $1.0329i$  \\
&1          & 1              & 0                      & 0.1                    & 0.3                   & $3.2054$ & $2.5081i$ & $1.9568$ & $1.7462i$ & $0.3574$ & $1.0407i$  \\
\hline
2&0.05          & 0.03              & 0                      & 0.1                    & 0                   & $2.8645$ & $2.2919i$ & $2.1797$ & $1.8750i$ & $0.2297$ & $1.0172i$  \\
&0.05          & 0.03              & 0                      & 0.1                    & 0.2                   & $3.0242$ & $2.3926i$ & $2.0553$ & $1.8026i$ & $0.3232$ & $1.0335i$  \\
&0.05          & 0.03              & 0                      & 0.1                    & 0.3                   & $3.0815$ & $2.4290i$ & $2.0102$ & $1.7767i$ & $0.3605$ & $1.0413i$  \\
\hline
3&0.05          & 0.03              & 0.001                      & 0.1                    & 0                   & $2.8640$ & $2.2921i$ & $2.1834$ & $1.8777i$ & $0.2292$ & $1.0181i$  \\
&0.05          & 0.03              & 0.001                      & 0.1                    & 0.2                   & $3.0242$ & $2.3931i$ & $2.0585$ & $1.8050i$ & $0.3230$ & $1.0344i$  \\
&0.05          & 0.03              & 0.001                      & 0.1                    & 0.3                   & $3.0817$ & $2.430i$ & $2.0133$ & $1.7791i$ & $0.3604$ & $1.0423i$  \\
\hline
4&0.05          & 0.03              & 0.001                      & 0.2                    & 0                   & $2.8634$ & $2.2916i$ & $2.1835$ & $1.8778i$ & $0.2279$ & $1.0178i$  \\
&0.05          & 0.03              & 0.001                      & 0.2                    & 0.2                   & $3.1302$ & $2.4604i$ & $1.9741$ & $1.7567i$ & $0.3935$ & $1.0497i$  \\
&0.05          & 0.03              & 0.001                      & 0.2                    & 0.3                   & $3.2118$ & $2.5125i$ & $1.9071$ & $1.7187i$ & $0.4533$ & $1.0646i$  \\
\hline
\end{tabular}}

\label{tab:lambda_col}
\end{center}
\end{table}

The stability of equilibrium points are studied by introducing the perturbation in the equilibrium point $(x_0,y_0)$, hence we define
\begin{align} \label{eq:perturb}
\begin{split}
    x&=x_0+\alpha, \\
    y&=y_0+\beta,
\end{split}
\end{align}
where $\alpha$ and $\beta$ is small displacements with respect to the equilibrium points. By substituting \cref{eq:perturb} to \cref{eq:eq_motion} and expand the equation, we get
\begin{align}
\begin{split}
    \Ddot{\alpha}&-2n\Dot{\beta}=W_{xx}^0\alpha+W_{xy}^0\beta, \\
    \Ddot{\beta}&+2n\Dot{\alpha}=W_{yx}^0\alpha+W_{yy}^0\beta,
\end{split}
\end{align}
where
\begin{align}
\footnotesize
\begin{split}
    W_{xx}&=n^2 + \frac{(1-\mu)q_1}{r_1^3}\left(-1+\frac{3x^2}{r_1^2}\right) + \frac{\mu q_2}{r_2^3}\left(-1+\frac{3(x-1)^2}{r_2^2}\right) + \frac{M_b}{(R^2+T^2)^{3/2}}\left(-1+\frac{3(x-\mu)^2}{(R^2+T^2)}\right), \\
    W_{yy}&=n^2 + \frac{(1-\mu)q_1}{r_1^3}\left(-1+\frac{3y^2}{r_1^2}\right) + \frac{\mu q_2}{r_2^3}\left(-1+\frac{3y^2}{r_2^2}\right) + \frac{M_b}{(R^2+T^2)^{3/2}}\left(-1+\frac{3(y-2k/n)^2}{(R^2+T^2)}\right), \\
    W_{xy}&=W_{yx}=\frac{3(1-\mu)q_1xy}{r_1^5} + \frac{3\mu q_2(x-1)y}{r_2^5} + \frac{3M_b(x-\mu)(y-2k/n)}{(R^2+T^2)^{5/2}}.
\end{split}
\end{align}
The characteristic equation is given by
\begin{equation}
    \lambda^4+\left(4n^2-W_{xx}^0-W_{yy}^0\right)\lambda^2+W_{xx}^0W_{yy}^0-\left(W_{xy}^0\right)^2=0.
\end{equation}
The solution of this equation is given as follows 
\begin{equation}
\lambda_i = \pm \sqrt{(-b \pm \sqrt{b^2 - 4c})/2}; \quad i=1,2,3,4.   
\end{equation}
where $b = 4n^2-W_{xx}^0-W_{yy}^0$ and $c = W_{xx}^0W_{yy}^0-(W_{xy}^0)^2$. The stability of equilibrium points can be achieved when all $\lambda_i$ are purely imaginary, otherwise we have an unstable equilibrium point. 

Table \ref{tab:lambda_col} shows the characteristic roots ($\lambda_i$) of the collinear equilibrium points by considering several configuration of perturbing parameters. All $\lambda_1$ have the form real which signify instability. In the range of mass parameter $0 < \mu < 1$ we found that $b^2-4c > 0$ for $L_1$, $L_2$, and $L_3$. Consequently we have at least one positive real for the solution of characteristic equation. Hence, the collinear equilibrium points are always unstable.

In the case of triangular equilibrium points, the stability is achieved when $0<\mu<\mu_c$, where $\mu_c$ means the critical mass. Following \cite{SinghTauraGCRTBP}, the critical mass is given as follows
\begin{align}
\begin{split}
    \mu_c&=\frac{1}{2}\left(1-\sqrt{\frac{23}{27}}\right)-2\frac{2-q_1-q_2}{27\sqrt{69}} \\
    &+\left(\frac{3}{2}+\frac{(76-8r_c)(r_c^2+T^2)}{27\sqrt{69}}-\frac{83+12r_c^2}{6\sqrt{69}}\right)\frac{M_b}{(r_c^2+T^2)^{5/2}}.
\end{split}
\end{align}
Table \ref{tab:lambda_tri} shows the examples of characteristic roots in the stability of triangular equilibrium points. All case have stable equilibrium points during $t=0$ and unstable in $t=0.2$ and $t=0.3$. We noted that $\lambda_1$ and $\lambda_3$ in $L_4$ are similar to $L_5$ for the case 1 and case 2. However, due to the combination of disk-like structure and mass transfer effects, this similarity is not sound for case 3 and case 4.

\begin{figure}
     \centering
     \begin{subfigure}[b]{0.32\textwidth}
         \centering
         \includegraphics[width=\textwidth]{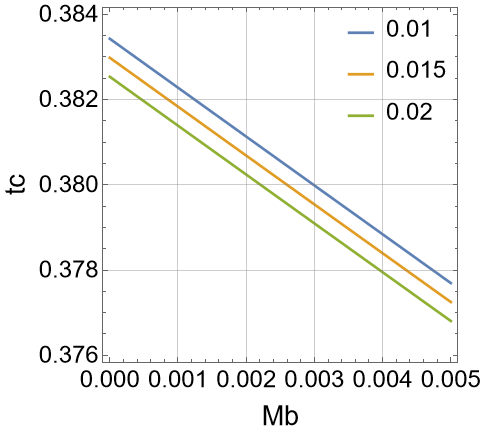}
         \caption{}
         \label{fig:tcMp}
     \end{subfigure}
     \hfill
     \begin{subfigure}[b]{0.32\textwidth}
         \centering
         \includegraphics[width=\textwidth]{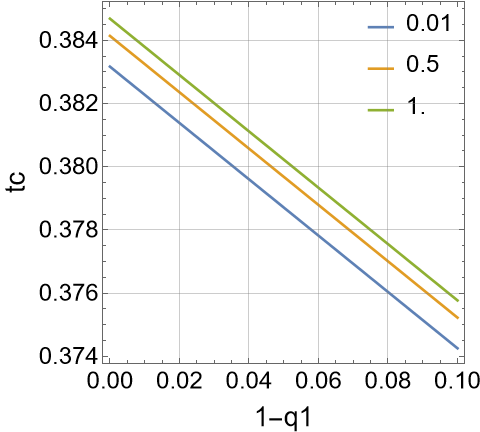}
         \caption{}
         \label{fig:tcpT}
     \end{subfigure}
     \hfill
     \begin{subfigure}[b]{0.32\textwidth}
         \centering
         \includegraphics[width=\textwidth]{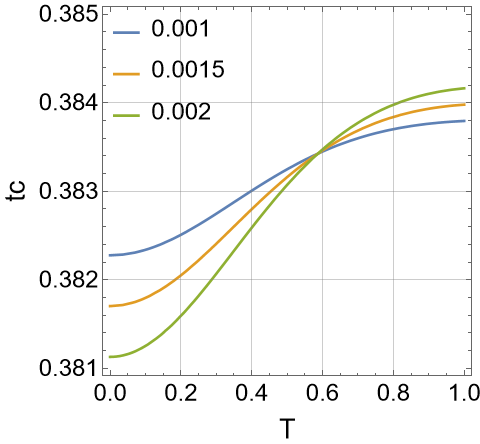}
         \caption{}
         \label{fig:tcTM}
     \end{subfigure}
        \caption{$t_c$ as a function of (a) $M_b$, (b) $q_1$, and (c) $T$, for various values of (a) $1-q_1$, (b) $T$, and (c) $M_b$.}
        \label{fig:tc}
\end{figure}

Since, in our case, $\mu$ depends on time, besides $\mu_c$ there exists also a so called critical time ($t_c$) as follows
\begin{equation}
    t_c=t_0+\left(\mu_c-\mu_0\right)/k,
\end{equation}
where $t>t_c$ means unstable. \cref{fig:tc} shows the effect of perturbing parameters $M_b$, $q_1$, and $T$ on $t_c$ for the case of $k=0.1$, $\mu_0=0.3$, and $t_0=3$. We noted that $t_c$ becomes shorter when $M_b$ and $1-q_1$ increase. In contrast, $t_c$ is longer if $T$ increases.

\begin{table}
\caption{Characteristic roots ($\lambda_{1}$ and $\lambda_{3}$) of triangular equilibrium points with $\mu = 0.02$. We used $T = 0.2$ and $t_0 = 0$. Here $i$ means $\sqrt{-1}$. $\lambda_{2}$ and $\lambda_{4}$ have the inverse sign of $\lambda_{1}$ and $\lambda_{3}$ respectively.}
\begin{center}
{\renewcommand{\arraystretch}{1.5}
\setlength{\tabcolsep}{4pt} 
\begin{tabular}{|c|c|c|c|c|c|c|c|c|c|}
\hline
\multirow{2}{*}{Case} & \multirow{2}{*}{$1-q_1$}   & \multirow{2}{*}{$1-q_2$}   & \multirow{2}{*}{$M_b$}   & \multirow{2}{*}{$k$} & \multirow{2}{*}{$t$} & \multicolumn{2}{|c|}{$L_4$}&\multicolumn{2}{|c|}{$L_5$} \\ 
\cline{7-8} \cline{9-10}
&&&&& & $\lambda_{1}$ & $\lambda_{3}$ & $\lambda_{1}$ & $\lambda_{3}$ \\
\hline
1 &1          & 1              & 0                      & 0.1                    & 0                   & $0.3961i$ & $0.9182i$ & $0.3961i$ & $0.9182i$  \\
 &1          & 1              & 0                      & 0.1                    & 0.2                   & $0.0675 + 0.7103i$ & $0.0675 - 0.7103i$ & $0.0675 + 0.7103i$ & $0.0675 - 0.7103i$   \\
 &1          & 1              & 0                      & 0.1                    & 0.3                   & $0.1820 + 0.7302i$ & $0.1820 - 0.7302i$ & $0.1820 + 0.7302i$ & $0.1820 - 0.7302i$  \\
\hline
2 &0.05          & 0.03              & 0                      & 0.1                    & 0                   & $0.4005i$ & $0.9163i$ & $0.4005i$ & $0.9163i$   \\
&0.05          & 0.03              & 0                      & 0.1                    & 0.2                   & $0.0828 + 0.7119i$ & $0.0828 - 0.7119i$ & $0.0828 + 0.7119i$ & $0.0828 - 0.7119i$   \\
&0.05          & 0.03              & 0                      & 0.1                    & 0.3                   & $0.1889 + 0.7319i$ & $0.1889 - 0.7319i$ & $0.1889 + 0.7319i$ & $0.1889 - 0.7319i$   \\
\hline
3&0.05          & 0.03              & 0.001                      & 0.1                    & 0                   & $0.4018i$ & $0.9167i$ & $0.4006i$ & $0.9174i$   \\
&0.05          & 0.03              & 0.001                      & 0.1                    & 0.2                   & $0.0835 + 0.7127i$ & $0.0835 - 0.7127i$ & $0.0822 + 0.7126i$ & $0.0822 - 0.7126i$   \\
&0.05          & 0.03              & 0.001                      & 0.1                    & 0.3                   & $0.1892 + 0.7326i$ & $0.1892 - 0.7326i$ & $0.1888 + 0.7326i$ & $0.1888 - 0.7326i$   \\
\hline
4&0.05          & 0.03              & 0.001                      & 0.2                    & 0                   & $0.4074i$ & $0.9139i$ & $0.4007i$ & $0.9174i$   \\
&0.05          & 0.03              & 0.001                      & 0.2                    & 0.2                   & $0.2499 + 0.7504i$ & $0.2499 - 0.7504i$ & $0.2477 + 0.750i$ & $0.2477 - 0.750i$   \\
&0.05          & 0.03              & 0.001                      & 0.2                    & 0.3                   & $0.3263 + 0.7792i$ & $0.3263 - 0.7792i$ & $0.3252 + 0.7790i$ & $0.3252 - 0.7790i$   \\
\hline
\end{tabular}}

\label{tab:lambda_tri}
\end{center}
\end{table}


\section{Conclusions}\label{sec6}

We have studied the influence of mass transfer, disk-like structure, and radiation pressure, on the position and stability of CRTBP equilibrium points. In this system, we found there are five equilibrium points, where two of them are triangular equilibrium points, and the others are quasi-collinear equlibrium points. Unlike the classical collinear equilibrium points, we noted that $L1$, $L2$, and $L3$ are slightly departed from x-axis since there exist the effects from disk-like structure and mass transfer. Moreover, the symmetry of $L_4$ and $L_5$ is broken when we consider the mass transfer and disk-like structure together. Furthermore, we found that the quasi-collinear equilibrium points remain unstable. The stability of triangular points depend on the initial mass parameter $\mu_0$ as well as the time. We found there exist critical time for achieving the stability of triangular points.

\section*{Acknowledgment}
We thank the reviewers for their insightful comments and suggestions on the manuscript.This research has been supported by RIIM LPDP-BRIN 2023-2025 and UI research grant No. PKS-026/UN2.F3.D/PPM.00.02/2023. 

\bibliography{sn-bibliography}

\end{document}